\documentclass[lettersize,journal]{IEEEtran}
\usepackage[utf8]{inputenc}
\usepackage{babel}
\usepackage{cite}
\usepackage{verbatim}
\usepackage{amsmath}
\usepackage{graphicx}
\usepackage{booktabs}
\usepackage{amssymb}
\usepackage{multicol}
\usepackage{flushend}
\usepackage{enumitem}
\usepackage{algorithm}
\usepackage{algpseudocode}
\usepackage[table,xcdraw]{xcolor}
\PassOptionsToPackage{hyphens}{url}
\usepackage[unicode=true, bookmarks=false, breaklinks=false,pdfborder={0 0 1}, colorlinks=false]{hyperref}
\usepackage{url}  % This makes \url work

\usepackage{float}
\newfloat{fn}{hb}

 \usepackage{tikz}
\newcommand*\circled[1]{\tikz[baseline=(char.base)]{
            \node[shape=circle,draw,inner sep=0.5pt] (char) {#1};}}

\setlist[itemize]{wide=\parindent}
\setlist[enumerate]{wide=\parindent}

\title{From Design to Deployment of Zero-touch\\Deep Reinforcement Learning WLANs}

\author{
\IEEEauthorblockN{
Ovidiu Iacoboaiea,~%\IEEEauthorrefmark{1}, 
Jonatan  Krolikowski,~%\IEEEauthorrefmark{1}, 
Zied Ben Houidi,~%\IEEEauthorrefmark{1}, 
Dario Rossi%\IEEEauthorrefmark{1}
}
    \IEEEauthorblockA{%\IEEEauthorrefmark{1}
    Huawei Technologies France SASU\\
    \texttt{\{ovidiu.iacoboaiea, jonatan.krolikowski, zied.ben.houidi, dario.rossi\}@huawei.com
}
}
}

\begin{document}
%to force authors et al. Must be the first command 
\bstctlcite{IEEEexample:BSTcontrol}

\maketitle
\begin{abstract} 
Machine learning (ML) is increasingly used to automate networking tasks, in a paradigm known as zero-touch network and service management (ZSM).   In particular, Deep Reinforcement Learning (DRL) techniques have recently gathered much attention for their ability to learn taking complex decisions in different fields. In the ZSM context, DRL is  an appealing candidate for  tasks such as dynamic resource allocation, that is generally formulated as hard optimization problems.  At the same time, successful training and deployment of DRL agents in real-world scenarios  faces a number of challenges that we outline and address in this paper. Tackling the case of Wireless Local Area Network (WLAN) radio resource management, we report guidelines that extend to other usecases and more general contexts. 
\end{abstract}

\section{Introduction}

WLAN has become the ubiquitous access technology at home, in public locations such as train stations, or private ones such as university or corporate campuses. Especially in densely populated areas, scarcity of radio resources can easily lead to congestion and thus bad user experience. Luckily, the  fleets of WLAN access points (APs) in campus networks can be centrally controlled,  opening the way for dynamic and autonomous configuration of network resources: as many of such dynamic resource allocation problems are hard, they are solved in practice using well-thought heuristics. 

Inspired by success of Machine Learning (ML), the field of communication networks has been actively seeking to exploit such techniques to automate complex network tasks, paving the way toward the realization of zero-touch network and service management. 
In particular, Deep Reinforcement Learning  (DRL) techniques, which learn by interacting with an environment, are able to  achieve complex tasks with unprecedented skills -- top stories include Google's AlphaGo~\cite{alphago} beating the Go world champion Lee Sedol in 2016, or OpenAI Five~\cite{openaiFive} winning an online computer-game DOTA2 tournament in 2017, 
or recent advances in fully autonomous cars from Tesla~\cite{TeslaPytorch}.  Following similar path, recent attempts to use DRL instead of heuristics  for automating network resource allocation~\cite{mao2016resource}, routing~\cite{valadarsky2017learning},
 WLANs configuration~\cite{IFIPnetworking} and more~\cite{Vesselinova_2020,zsmieeenet20,zsm-wlan22} have shown promising results. 
At the same time, we observe that while it is relatively straightforward to design and train DRL agents that work well in synthetic and controlled settings, \emph{real-world deployment} of the same DRL agents poses a set of additional challenges. Indeed,  performance evaluation in simplified settings remains a \emph{necessary} task (i.e., if a solution does not work in simulation, it will never work in the real world), but it is clearly not  \emph{sufficient}  (i.e., there are no guarantees that the DRL solution will work as expected in a different environment than the ones on which it was trained). Thus, in order to carry DRL all the way from design to deployment, a number of practical and often underestimated challenges must be accounted for. The latter are just as important as the ML algorithmic design. 

Such challenges are rooted in the architectural requirements that must be fulfilled  in order for ML techniques to be seamlessly applied in the network: these are nailed down by standardization bodies, as for instance ETSI Zero-touch Network and Service Management (ZSM)\footnote{\begin{scriptsize}\url{https://www.etsi.org/technologies/zero-touch-network-service-management}\end{scriptsize} accessed on 01.07.2022}, that  provide normative architectural references for several tasks.
Clearly, as ML techniques are data-driven, a set of requirements concern access to telemetry data, notably the ability to  stream it (ZSM requirement \#84),  enforce access control (\#86) and store it in data lakes (\#87). In particular, DRL model training  requires access to data lakes (and likely GPU resources), whereas DRL model inference requires access to stream telemetry (and significantly less computational resources).
Furthermore,  ML operation requires the ability to deploy and upgrade trained ML models  (\#46, \#49):  while some ``default'' trained model may be necessary for generic zero-touch operation, the same model may be ``fine-tuned'' to the specifics of the environment after deployment. Training a generic model requires  historical data gathered from several networks and available in a data lake, while model upgrading requires fresh telemetry for the purpose of fine-tuning.  Finally, and most importantly, 
closed-loop techniques such as DRL need the ability to enforce actions automatically (\#68, \#115), depending on specific conditions determined by the algorithm, in order to adapt resources allocation better to the instantaneous or forecast evolution of service load.

This paper reports our experience in designing and deploying DRL for zero-touch WLAN networks.  We build over our original design of a  DRL sequence-to-sequence architecture, 
that we  limitedly validated for the purpose of WLAN resource allocation  in simulated settings~\cite{IFIPnetworking}, and that have now been running for months on real operational deployments. In the path from design to deployment, we outline and tackle five important challenges related to  (i)  safety, (ii) duration and (iii) realism  of the training process,  as well as the (iv) generalization capabilities and (v) the adoption barrier  of   trained models. In sharing our experience with the community, we not only illustrate the specific way in which we overcome such challenges in the WLAN case, but further adopt a broader viewpoint: we   complement lessons learned with those gathered from other fields where DRL has been successful, such as gaming\cite{alphago,openaiFive} and self-driving cars\cite{TeslaPytorch,AWSdeepracer},  testifying the generality of these challenges.

The rest of the paper presents a high-level view of the zero-touch WLAN resource management problem (\S\ref{sec:bg}), articulates the main challenges from design to deployment (\S\ref{sec:wlan:practice}), and summarizes the main lessons (\S\ref{sec:conclusion}).

\section{Zero-Touch Deep Reinforcement WLANs}\label{sec:bg}
Our goal is to autonomously manage WLANs in closed-loop, continuously adapting allocated radio resources to changing traffic conditions and demand, to maximize end-to-end performance.  We first briefly cover WLAN management (\S\ref{sec:wlan:bg}), that we next reconsider under the lens of zero-touch closed-loop control   (\S\ref{sec:drl:zerotouch}),
and finally overview the DRL technique we employ
(\S\ref{sec:drl:bg}).

\subsection{Wireless LANs}\label{sec:wlan:bg}
WLANs are defined in the IEEE 802.11 standard\footnote{\url{https://www.ieee802.org/11/} accessed on 01.07.2022} and its amendments. Zero-touch operation in the general case \cite{zsmieeenet20} and in  heterogeneous, industrial and enterprise WLANs are surveyed in~\cite{zsm-wlan22}. Here we provide a very basic overview of WLAN resources and actions from the viewpoint of autonomous closed-loop control.  The most popular WLAN setup is infrastructure-based, where  stations (such as smartphones, laptops or industrial devices, referred to as STAs) connect to  fixed Access Points (APs) that typically act as gateways to relay STA traffic. 
In office buildings or university campuses, a fleet of APs is deployed over a large area to connect the numerous STAs to the Internet. Typically,  centralized management decisions are taken to  optimally manage the network: the set of actions include\footnote{But are not limited to: e.g., consider low-level configuration parameters related to antenna parameters, MIMO, backoff timers, etc.}, for each AP, selecting a channel, bonding and power configuration.  

 Each AP is configured to use a specific \emph{primary channel} (within the 2.4GHZ or the 5GHz band), performing downlink and uplink transmissions in a half-duplex manner.  
 Optionally, an AP may be configured to allow the aggregation (aka \emph{bonding}) of several channels, to increase bandwidth and consequently throughput.
Ideally, only one device within receiver vicinity transmits on one channel at the same time, avoiding collisions and data loss.  This is achieved through the listen-before-talk mechanism of carrier-sense multiple access with collision avoidance (CSMA/CA). As a consequence, APs and STAs on the same channel share airtime: the time that a transmitter  waits  while the channel is busy is called interference (time). 
Depending on the regulatory region, only 4 (20) channels are non-overlapping on the 2.4\,GHz (5\,GHz) band: thus, it is not always possible to allocate different channels to neighboring APs and, in highly dense areas,  interference cannot be avoided by simple channel allocation. It follows that on top of channel allocation and bonding~\cite{TurboCA},  also the AP  {\em transmit power}~\cite{Keshav,  power_limitations} can be additionally used to tradeoff the strength and quality of the received signal vs the airtime interference.

\begin{figure}[t]
\centering
\includegraphics[width=1\columnwidth]{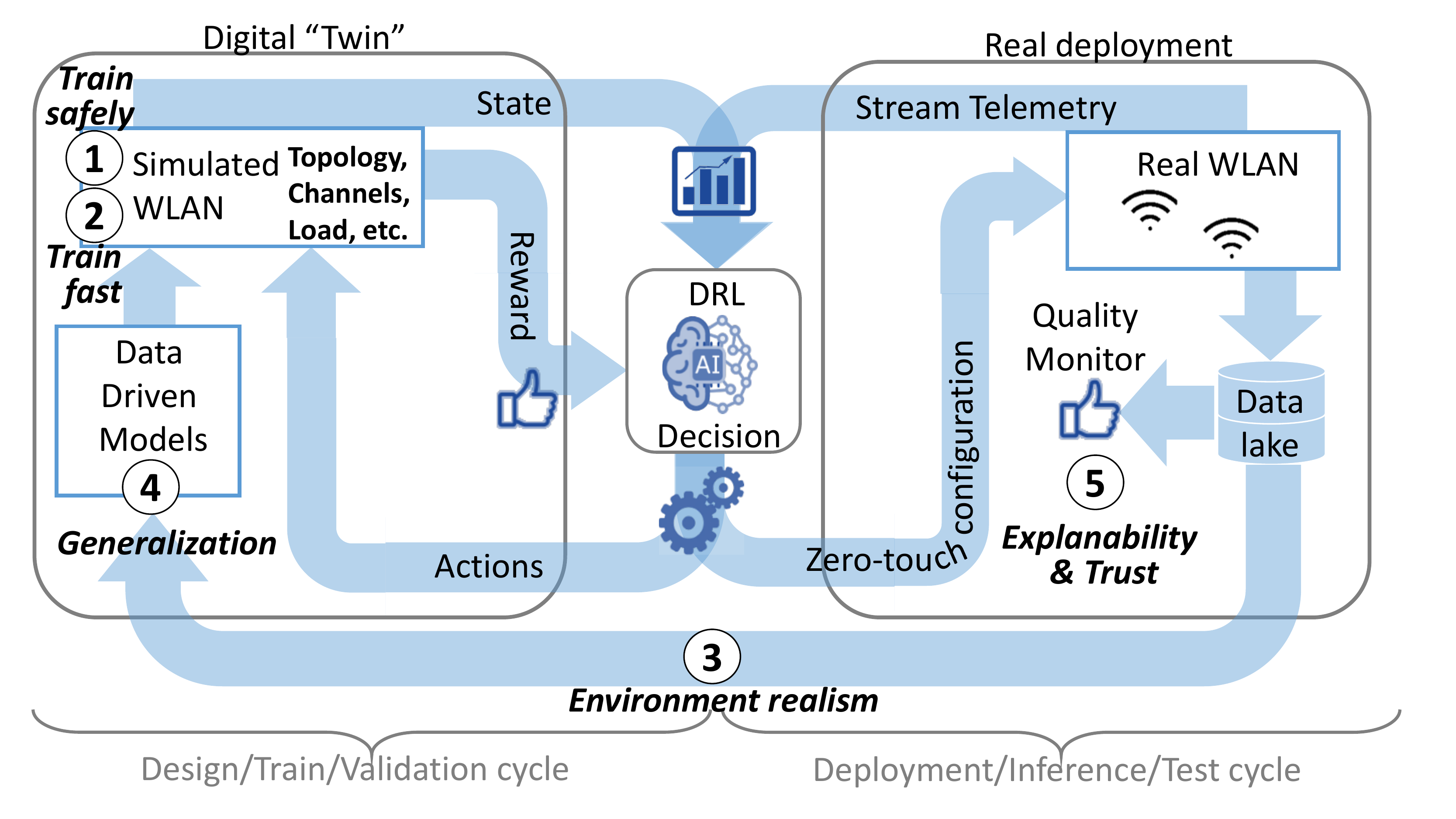}
\caption{Zero-touch WLANs loops:  design/train/validation using a digital replica of the real network (left)  and deployment/inference/test in the operational network (right).  }\label{fig:wlan_overview}
\end{figure}

\subsection{Zero-touch WLANs}\label{sec:drl:zerotouch}
We note that the network configuration has to be selected from a very large state space, that grows exponentially with the set of available configuration knobs.  Additionally, as the network load evolves over time, it would be desirable for the WLAN to be able to autonomously adjust its configuration to best adapt  the available resource to the current (or forecast future) demand.
%%%%
As the utility function to estimate network quality can be a complex combination of QoS (e.g., signal strength, coverage, interference, user throughput, latency) and QoE indicators (e.g. more advanced per-application metrics), this makes autonomous configuration a desirable capability of WLAN networks. From this viewpoint, with reference to Fig.~\ref{fig:wlan_overview}, it is envisionable that a zero-touch WLAN is governed by an ML model taking decisions (e.g. configuration actions) as a function of observable state (e.g. stream telemetry). Such a model should be pre-trained (e.g., by using a digital replica\footnote{Such a digital replica is commonly referred to as \emph{digital twin}; however, twins are very faithful representations while, as we shall see, an approximated replica suffices for our training requirements, hence the quotes.} of the network%, for immediate action
), but could possibly benefit from specific fine-tuning from real-data after deployment (to upgrade the model in the long run).   

We observe that the existence of two separate environments results in a  dichotomy  inducing three separate loops: a  first design/train/validation cycle (left of Fig.~\ref{fig:wlan_overview}) where the ML model is trained on a digital replica of the network;  a second deploy/inference/test cycle (right of Fig.~\ref{fig:wlan_overview}) where the trained model is used on the actual network; a third refinement loop, bridging the two environments.  
The picture also highlights several important practical aspects that this paper is going to dissect, notably: pre-training is necessary to \circled{1} {\it train safely} and \circled{2} {\it train fast}, but note the need for \circled{3} {\it environmental realism} for better fit   and \circled{4} {\it  generalization capabilities} to unknown states. Finally, \circled{5} {\it explainability and trust} are key to deploy zero-touch closed-loop operation.

\subsection{Deep Reinforcement WLANs}\label{sec:drl:bg}

DRL techniques are suitable for implementing the closed-loop control algorithm.  
To better understand challenges that DRL agents may face in real-world deployment, we first briefly remind its most important concepts, and cast them to WLAN with the help of Fig.~\ref{fig:wlan_overview}.
Without loss of generality, in the the context of this work  we limit the configuration knobs to the  the selection of primary channel and bonding. 

\begin{figure}[t]
\centering
\includegraphics[width=1\columnwidth]{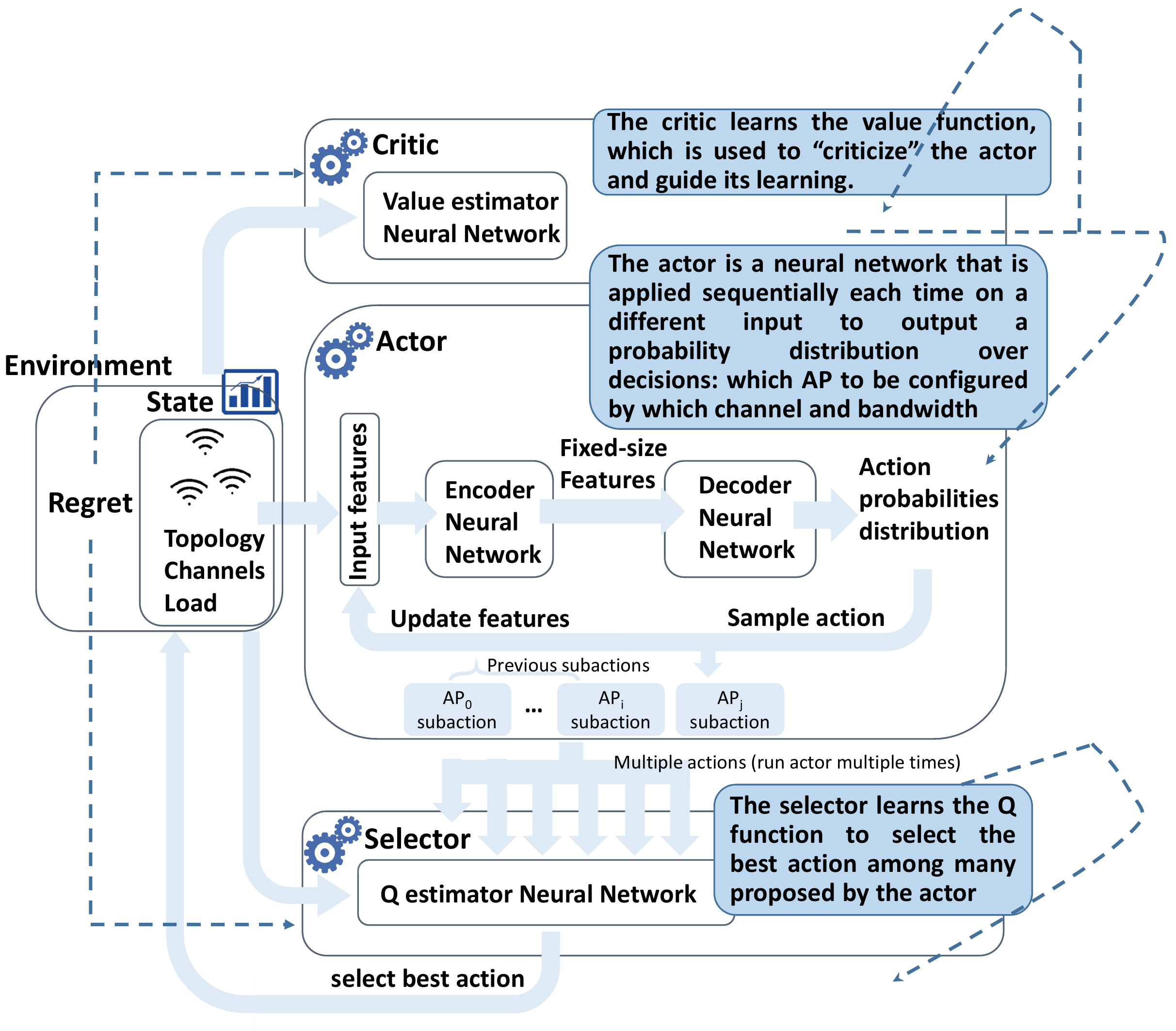}
\caption{Simplified synoptic of the DRL architecture for  WLAN management, described in more details in our prior work~\cite{IFIPnetworking}.}\label{fig:DRL_overview}
\end{figure}

\subsubsection{Background on RL and DRL}
In  Reinforcement Learning (RL)~\cite{Sutton1998}, an \emph{agent} learns from interacting with an \emph{environment}: the agent obtains a perception of the environment through a  measurable \emph{state}  (e.g.,  network configuration, stream telemetry, etc.) and selects an \emph{action}  (e.g., changing an AP configuration) based on a policy that it is learning. After enforcing the action, the agent observes an updated state and receives feedback about its action, in the form of a \emph{reward}  (or a \emph{regret}). Unlike in supervised learning (where the feedback reflects some distance from the optimum),  the feedback in RL can be rather seen as praise (or critique) of the action (without any explicit information about the optimality).  Based on this feedback, the agent updates its policy, observes the new environment state and enforces a new action, learning to increase its reward (or decrease its regret).
Deep Reinforcement Learning (DRL) is a class of approaches that are  based on a Neural Network (NN), where the NN is used  either to learn the value of the state (i.e. DQN) or the policy (e.g. A2C, A3C). A recent wave of 
DRL approaches have shown interesting results in the solution of combinatorial graph problems, using various architectures, such as Graph Neural Networks, pointer networks and graph attention networks~\cite{Vesselinova_2020}.

\subsubsection{Overview of WLAN DRL agent} 
In our previous work~\cite{IFIPnetworking}, we used a similar philosophy to develop a DRL architecture fit for WLAN channel management.   
While in~\cite{IFIPnetworking} we limitedly validate the approach against state of the art via \emph{simulation} (left of Fig.~\ref{fig:wlan_overview}), in this paper we are concerned about the complementary necessary steps for  \emph{real-world deployment} (bridging the gap between left and right of Fig.~\ref{fig:wlan_overview}). 
As such, we provide here only a necessary limited overview of the WLAN DRL in Fig.~\ref{fig:DRL_overview}, and refer the reader to ~\cite{IFIPnetworking} for details.  

Our design follows a classic actor-critic Neural Network (NN) architecture, where the critic-NN which learns the value function guides the actor-NN in learning the best policy, to which we add a selector-NN, guiding the choice of the best action among those returned by multiple parallel runs of the actor-NN. 
The actor-NN employs an encoder-decoder sequential architecture, where basically the same NN is run sequentially multiple times, picking at each step a decision for one of the APs  in the network.
In particular, our encoder-NN employs a careful feature engineering process to transform variable-size input features (which depend on network size, number of channels etc.) into a fixed-size intermediate representation, that is used as input by the decoder-NN to output a probability distribution over all actions (i.e. one channel and bandwidth option for each AP), which makes it suitable for application to arbitrary networks.

\section{From design to deployment}\label{sec:wlan:practice}

We now report on our deployment experience of DRL agents in real operational WLANs.  In particular, our DRL-based WLAN channel management solution ({\it agent})  autonomously reconfigures in closed-loop  ({\it action})  every 10 minutes (so that the need for accurate forecast of future demand is lessened by  the fact that actions are frequently taken)  a real operational WLAN ({\it environment}), based on telemetry data ({\it state}) received at sub-minute timescale. 

While training and deploying this DRL-based system, we faced a series of challenges \circled{1}-\circled{5} mentioned in Fig.~\ref{fig:wlan_overview},  into which we now dig deeper -- first summarizing our experience with WLAN deployment, and next contrasting the lessons learned to other DRL real-world use-cases.

\subsection{Train safely}\label{result:safety}
\subsubsection{WLAN insights}
Training on the real WLAN network would inevitably lead to the exploration of bad network configurations harming user experience. As this option is not viable for business considerations, and while an expensive WLAN testbed is not available, we are forced to train the DRL agent using a
digital replica of the system, such as a  computer-\emph{simulated} model of the real environment. While learning from a digital ``twin'' solves the  safety concerns altogether, it does however introduce another tradeoff.
Namely,  the simulator needs to be  \emph{realistic enough} to favor the transfer of learning to the real deployment  (\S\ref{result:data}) and at the same time   \emph{simple enough} to allow for reasonable training time (\S\ref{result:time}), which are both key aspects that are worth digging into deeper.

%\subsubsection{Lesson learned and outlook} 
\subsubsection{Beyond WLAN} 
In autonomous driving, the necessity to  training safely is even more obvious. For example, AWS deepracer\cite{AWSdeepracer}, a platform to train and test  DRL agent for this usecase, relies on a cloud based 3D racing simulator as one key component that helps avoiding the exploration of the most detrimental states even with model cars. Tesla also trains its Autopilot offline before allowing it on the road \cite{TeslaPytorch}.

\begin{figure}[t]
\centering
\includegraphics[trim=0.5cm 0.5cm 1.0cm 1.5cm, width=0.5\textwidth]{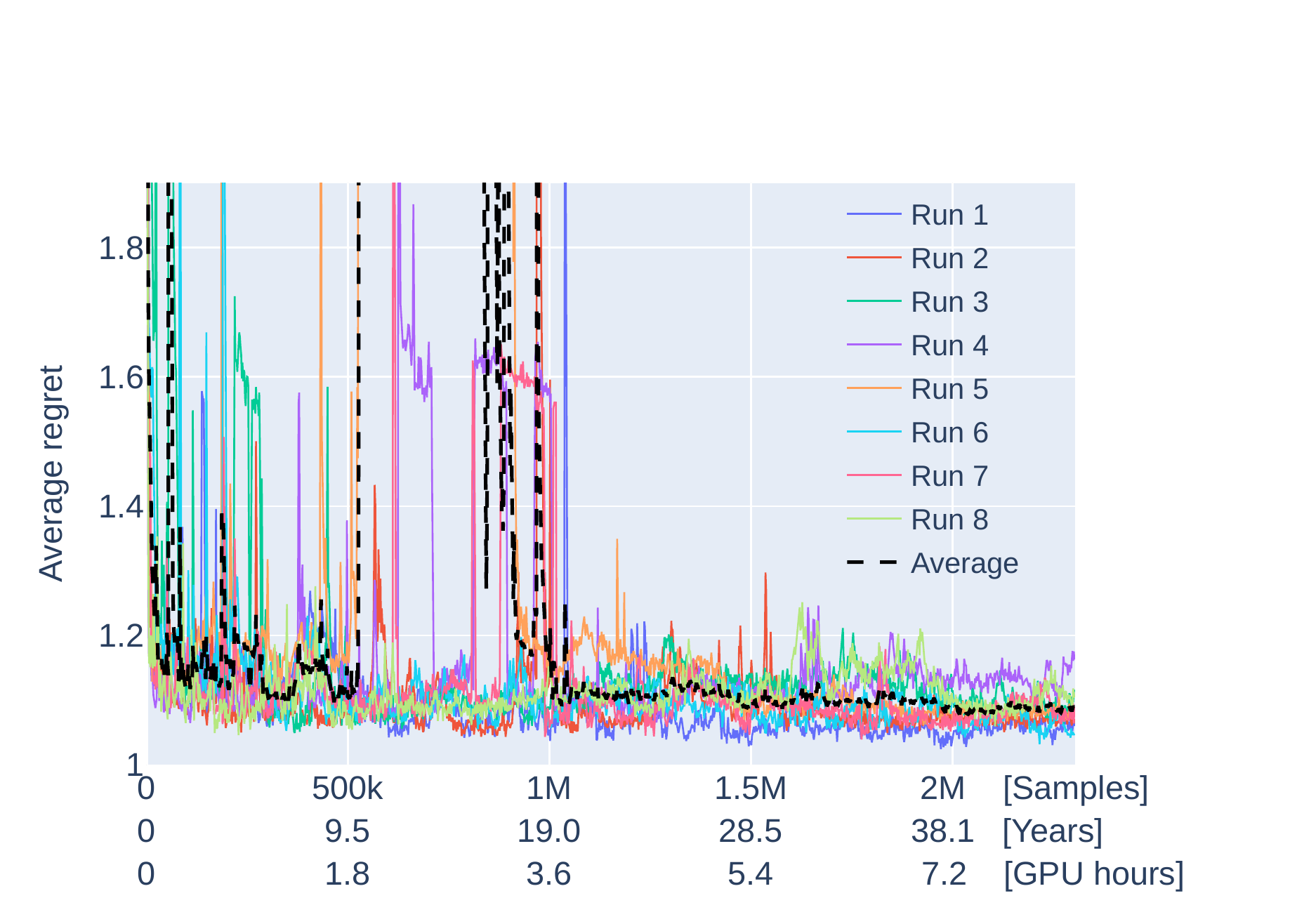}
\caption{\emph{Train fast}. Average regret (dashed black line) during training converges after more than  1 million interactions with the environment: this would require several years in a real deployment (at the considered timescale) and already requires about 8 hours worth of GPU time.} 
\label{fig:tensorboard5}
\end{figure}

\subsection{Train fast}\label{result:time}
\subsubsection{WLAN insights}
Two factors impact the training duration of our agent: (i) the convergence of DRL weights during the training process, which affects the number of interactions, and (ii) the duration of  each simulated interaction, during which the DRL training process remains idle waiting to receive state and regret feedback from the environment.
As for (i), we calibrated the training phase carefully to avoid getting stuck in local minima, mainly by adjusting the learning rate during training such that more aggressive updates (e.g. larger steps) are performed at the beginning, followed by smaller steps allowing the system to gradually stabilize. 
As for (ii), the duration of a simulated interaction can quickly become a bottleneck, for which we rule out the use of packet-level simulations (such as ns-2 or ns-3) and leverage a fast custom  simulator, with low computational complexity.
Fig.~\ref{fig:tensorboard5} illustrates the regret evolution over multiple independent training runs: the x-axis reports the number of iterations, the GPU training time (including the simulation time, measured in \emph{hours}) and the equivalent duration of the training process had it been performed at the same timescale in a real environment (measured in \emph{years}).

% \subsubsection{Lesson learned and outlook}
\subsubsection{Beyond WLAN} 
Training duration is a clear bottleneck in any DRL deployment. 
In the most recent successful DRL applications,  agents need several ``lifetimes'' of interaction with the environment, e.g.\ the 10,000 years equivalent of gameplay  for OpenAI Five~\cite{openaiFive}, which is clearly  unrealistic for training on real systems.
Even training offline with  real data  can take a significant amount of time: 
for instance, it takes 70,000 GPU hours to train the full self-driving Tesla pilot prototype~\cite{TeslaPytorch}, which is around one year for a single node with 8 GPUs. In some cases, training needs to be offloaded to a large fleet of data center servers equipped with  GPUs and TPUs, which may be only affordable for a few big players. in network usecases, depending on the size of the DRL NN, the digital "twin" can become the computational bottleneck.

\begin{figure}[t]
\centering
\includegraphics[%trim=0.0cm 1.0cm 0.0cm 1.5cm, 
width=0.5\textwidth]
{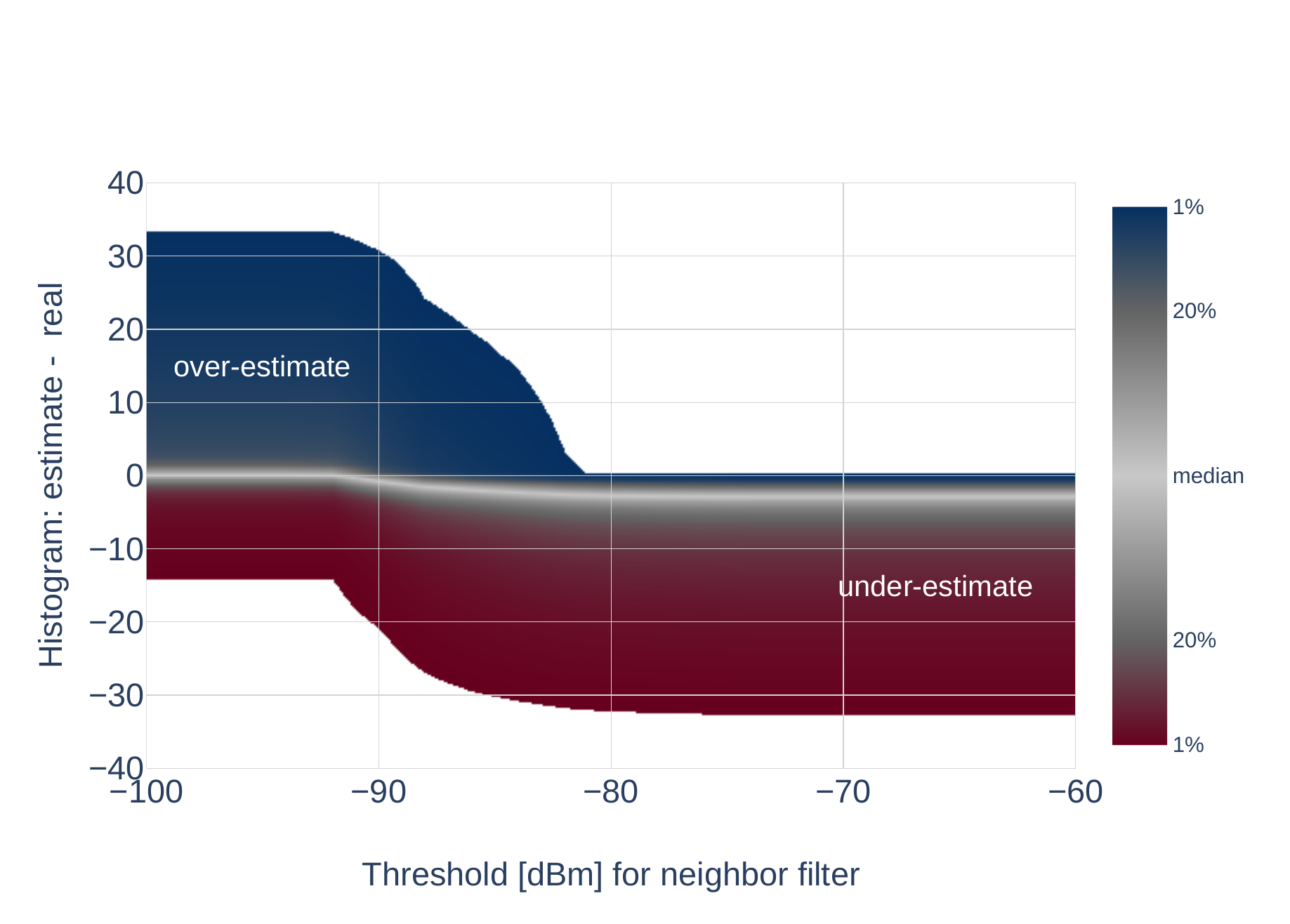}
\caption{\emph{Environment realism}. Augmenting the simulator models with data from real environment (fitting the interference threshold according to real neighborhood data).} 
\label{fig:thresholdcalibration}
\end{figure}

\subsection{Environment realism}\label{result:data}
 
\subsubsection{WLAN insights}
As environment realism remains a key concern, we can leverage real-world data to enhance the simulator models -- briding the training and validation environments. To retain scalability while enhancing realism, we use data-driven approach to refine several models used in the simulator.
%%%
To make just a single example in reason of space limits, our simulator uses a RSSI threshold to decide which APs are considered as neighbors: we increase realism by  fitting this parameter to maximize similarity between the interference estimations of the simulator and those measured in the real network. As can be observed in Fig.~\ref{fig:thresholdcalibration}, the ideal RSSI threshold under which two APs should be considered neighbors is at -82dBm, with which we calibrate the simulator for fine-tuning the DRL agent.

%\subsubsection{Lesson learned and outlook}
\subsubsection{Beyond WLAN} 
Environment realism clearly is key when an agent trained on simulation is to be transferred into the real world. When training on a simulator, sufficient resources need to be invested into its calibration~\cite{AWSdeepracer}.  We point out that alternatives to simulation exist, which may however not be fit to network use-cases. For instance, \emph{imitation learning} is a valid complementary  approach,  where instead of interacting with a real or simulated environment,  a database of state-action traces is used   for  offline batch-based agent training. For instance,
thanks to its  fleet of several hundred thousand self-driving-ready cars, Tesla  now has a  huge amount of state-action pairs (over 1 billion miles with Autopilot-on~\cite{TeslaPytorch}), which might be used to learn to  mimic human driver behavior.

However,  data-driven approaches in general, and in networks in particular, may be  vulnerable to data scarcity and quality issues. For instance, while in WLAN we have access to live measurements and long historical data from real networks, since the network is sparsely reconfigured (once per day in legacy data), our datalake is limited in terms of number of explored states, which DRL would require for training. Additionally, the states explored currently (i.e., network configurations) are limited to the subset induced by the existing algorithm, further restricting the boundaries of what DRL could possibly learn. Thus, simulations, augmented with data-driven models, are a better option.

\begin{figure}[t]
\centering
\includegraphics[width=0.47\textwidth]{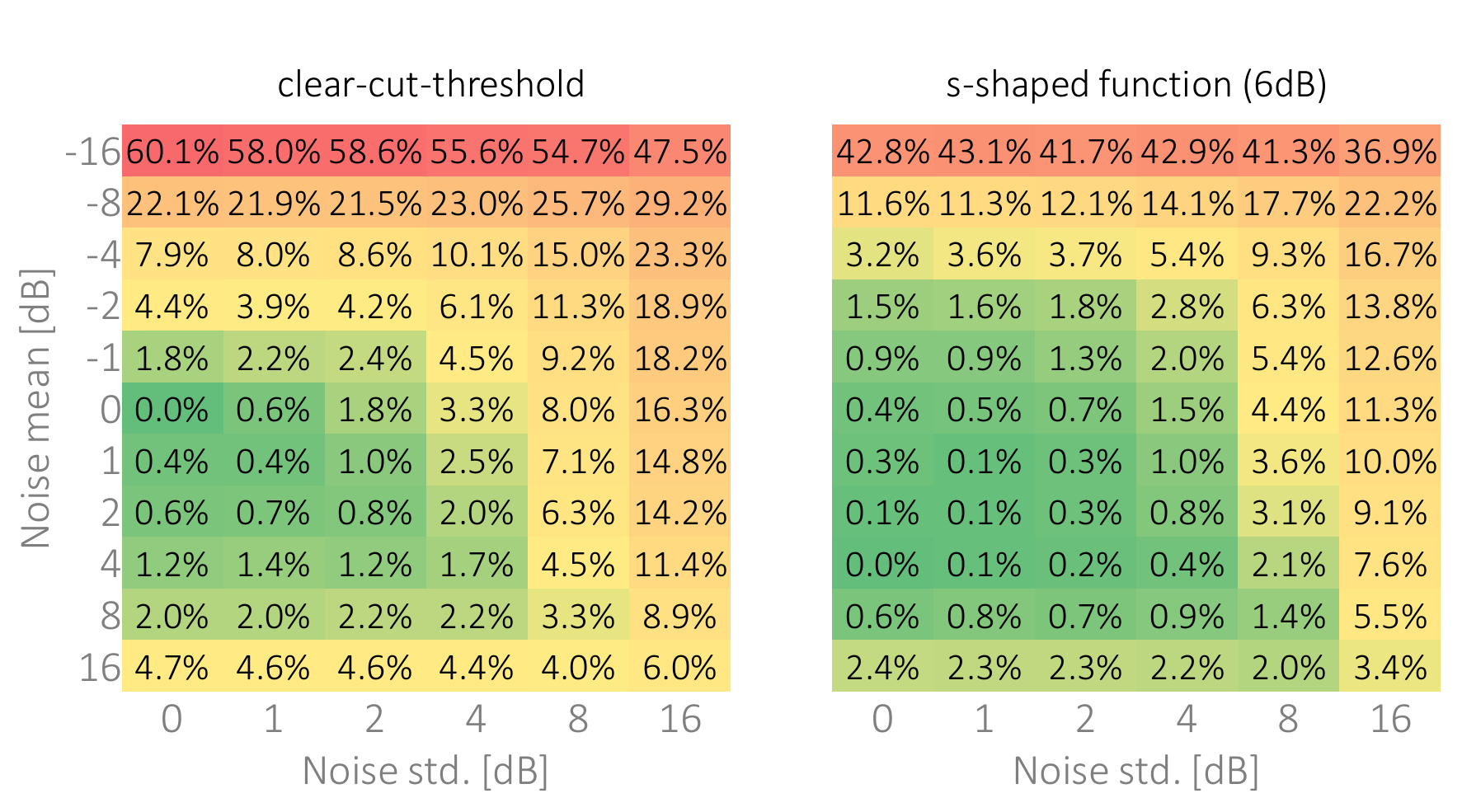}
\caption{\emph{Generalization}: Assessment  of relative performance degradation to controlled environmental changes by exposing the DRL agent to states distribution different than the one explored at training time.
}\label{fig:rb_all}
\end{figure}

\subsection{Generalization}\label{result:general}
\subsubsection{WLAN insights}
Generalization to conditions unseen during training means, on the one hand (i) generalizing to WLAN networks of  arbitrary size and density, and on the other hand (ii) transferring well to the more complex physics of the real network -- which are both necessary as deployment conditions will never match exactly the training conditions.
We tackle (i) by a novel auto-regressive sequential decoder whose input features at each step are engineered to reflect the changing internal state of the decoder that is described in \cite{IFIPnetworking}.

As for (ii), robustness of the DRL agent to varying conditions is key: to test the ability of the trained DRL model to  cope with unknown conditions, we train the agent in an ideal environment and test it in a noisy one. We systematically apply  Gaussian noise with controlled  means and standard deviations to the AP neighborhood (i.e., RSSIs each AP sees from all others) and observe the impact on the regret.  The left-hand side of Fig.~\ref{fig:rb_all} reports the relative percentual increase of the regret with respect to ideal conditions (noiseless training and testing) when neighborhood is defined with a simple threshold. Overall, the picture confirms DRL to be  robust for a wide range of additive noise. Additionally, consider that 
a negative noise causes the corresponding RSSI to fall below the neighborhood threshold, leading to interference underestimation. Unsurprisingly, the figure confirms underestimation to be more harmful than overestimation, which validates our conservative simulator design choice. 

However, in the real network, neighborhood is not exactly a clear-cut threshold:
the right-hand side of Fig.~\ref{fig:rb_all} further tests the algorithm on the same noisy conditions, but using a more complex neighborhood definition: in particular, neighborhood interference is smoothly taken into account by using an S-shaped sigmoid function with a spread of 6dB centered around the clear-cut threshold. Interestingly, using this ``loose'' definition does not degrade the results and leads to even better resistance to noise, suggesting good generalization ability.  

%\subsubsection{Lesson learned and outlook}
\subsubsection{Beyond WLAN} 
 
In any DRL deployment, when digital twins  are used for training, realism of the simulation is of primary concern --  the less faithful the simulator, the lower  the quality of the agent. In our case, we analyze the response to the trained model by stress-testing it against environmental changes.
We point out that (i) synthetic noise  or (ii) real-world data can be readily incorporated during the training process, although the exact way to do so depend on the specific usecase.

\begin{figure*}[t]
\centering
\includegraphics[%trim=0.5cm 1.0cm 1.0cm 1.5cm,
width=1\textwidth]{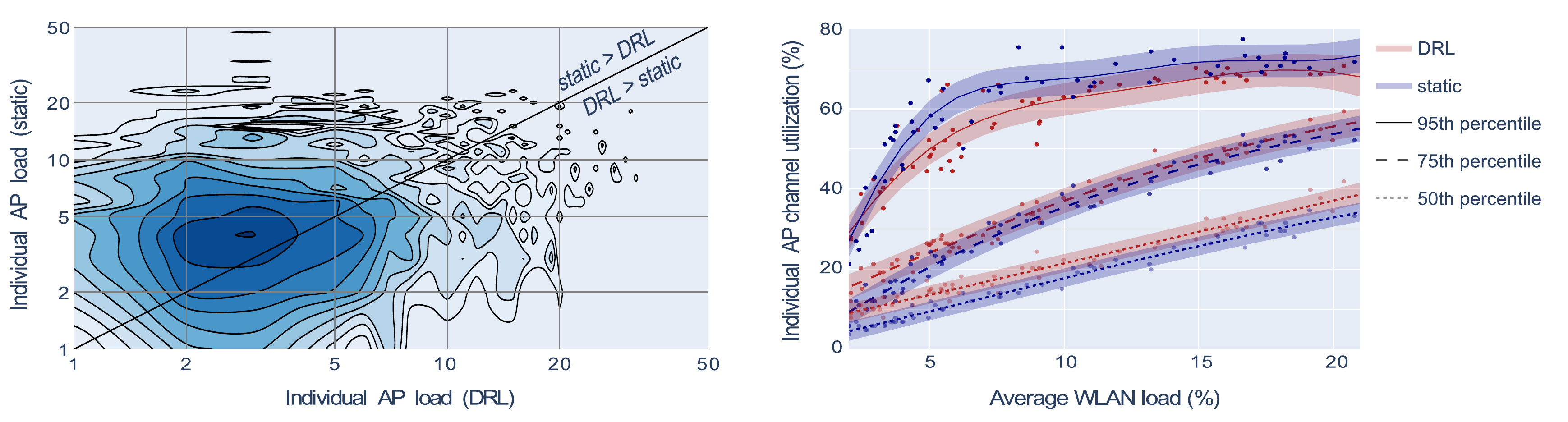}
\caption{\emph{Explainability (Left)}:  Channel utilization heatmap ,  comparing same AP  and same time-of-day slots across different algorithms (static and DRL) over  different days. \emph{Trust (Right)}: Statistically unbiased comparison of individual AP load (y-axis) for same average network load (x-axis). }
\label{fig:explain_and_trust}
\end{figure*}
 
\subsection{Explainability and Trust}\label{result:explain}  

\subsubsection{WLAN insights}

Lastly, it is essential that network engineers develop trust in the algorithm decisions before letting it run unattended on thousands of customer deployments. 
As DRL decisions are intrinsically less interpretable than heuristics,  this can firstly be achieved by human-understandable explanations of the algorithm decisions and expected gains --  so that the WLAN operator can not only perceive DRL operation as safe,  but also understand and value  its benefits.

Additionally, we conducted months worth of tests in operational WLAN networks to illustrate the DRL agent viability. In particular, we ran several batches of experiments, each lasting 1 week, in which we either run the (i)  trained  DRL agent every 10 minutes to closely track load changes vs (ii) a  daily static optimization based on historical load forecast. Here, we show an except of  deployment results in a Campus network in Nanjing, China: Fig.~\ref{fig:explain_and_trust} (left)  contrasts the channel utilization on a 30\,AP WLAN where we observe several thousand STAs on a typical day.
We construct a heatmap from the scatter plot where each point represents the average channel utilization for the same AP during 10 minutes at the same time-of-day and day-of-week  for the two algorithms, over all APs: this allows to assess the impact of dynamic DRL channel management from a spatial viewpoint, i.e. from the point of view of the same AP.  We can note the tendency of improvements as the \emph{center} for the highest density moves \emph{above} the diagonal.

Clearly, while the traffic is similar in every week due to seasonal behavior of the users, the traffic conditions are not identical, which can bias the comparison. Trust in the solution can be then gained only over a careful analysis over long-term campaigns. For instance,  we take this confounding factor into account  by comparing in Fig.~\ref{fig:explain_and_trust} (right) the breakdown of the AP utilization (y-axis) for the same average network load (x-axis): it is easy to see that, as expected, DRL relieves APs with high channel utilization (notice the 95th percentile decrease) by shifting load to lightly loaded APs (notice the median increase), which is desirable from the perspectives of load balancing and fairness. Overall, thorough testing under  real-world  conditions delivers a strong argument to IT engineers to trust the DRL agent for network O\&M.

%\subsubsection{Lesson learned and outlook}
\subsubsection{Beyond WLAN} 

As a general lesson,  human operators need to gain understanding and  build trust in  DRL systems to allow their deployment. Explainability of the algorithm output helps lowering the adoption barrier. Trust may then be gained step-by-step: convincing results from extended experiments on real-world deployment help showing the benefits of the algorithm -- which holds for any use-case. 

For instance, Google DeepMind adopted the following strategy for data center cooling: the first RL algorithm version\footnote{\url{https://deepmind.com/blog/article/deepmind-ai-reduces-google-data-centre-cooling-bill-40} accessed on 01.07.2022} acted only as a recommendation engine for the operator. Only later they moved to a fully autonomous version\footnote{\url{https://deepmind.com/blog/article/safety-first-ai-autonomous-data-centre-cooling-and-industrial-control} accessed on 01.07.2022}, still maintaining a failsafe option to revert back to human control at any time, in addition to rule-based heuristics as backup.

\section{Conclusion}\label{sec:conclusion}

 DRL is a promising paradigm for controlling complex systems,  improving decision making over human intuition and classic heuristics. While most network-related DRL research  focuses on ideal scenarios and are evaluated via simulation, we discuss here the challenges that arise when DRL is deployed in  large-scale operational WLAN to achieve zero-touch operation. 
 
 We generalize and summarize the lessons learned as follows.  It appears that DRL training requires digital ``twins'', such as simulators. Indeed, solely learning from existing network data may not be feasible (given the sheer number of samples needed for training) nor desirable (as it does not offer enough action diversity, so missing unsafe actions).
Conversely, learning from simulation provides the best tradeoff among safety (i.e. to  explore also unsafe actions),  simplicity (for training duration) and  realism (e.g.  as simulators can be enhanced  with real-world data and be used to assess controlled generalization). 
Finally, technical benefits are a necessary (but not sufficient) condition to adoption: deployment of trained DRL models for real-time  inference still requires a pedagogic effort toward the human operators interacting with it (in order to gain their trust), as well as offering fallbacks to legacy systems until  the algorithm gains sufficient trust for true fully-automated zero-touch operation.

\bibliographystyle{IEEEtran}
\bibliography{shortbib}

\begin{thebibliography}{10}
\providecommand{\url}[1]{#1}
\csname url@samestyle\endcsname
\providecommand{\newblock}{\relax}
\providecommand{\bibinfo}[2]{#2}
\providecommand{\BIBentrySTDinterwordspacing}{\spaceskip=0pt\relax}
\providecommand{\BIBentryALTinterwordstretchfactor}{4}
\providecommand{\BIBentryALTinterwordspacing}{\spaceskip=\fontdimen2\font plus
\BIBentryALTinterwordstretchfactor\fontdimen3\font minus
  \fontdimen4\font\relax}
\providecommand{\BIBforeignlanguage}[2]{{%
\expandafter\ifx\csname l@#1\endcsname\relax
\typeout{** WARNING: IEEEtran.bst: No hyphenation pattern has been}%
\typeout{** loaded for the language `#1'. Using the pattern for}%
\typeout{** the default language instead.}%
\else
\language=\csname l@#1\endcsname
\fi
#2}}
\providecommand{\BIBdecl}{\relax}
\BIBdecl

\bibitem{alphago}
D.~Silver, J.~Schrittwieser, K.~Simonyan, I.~Antonoglou, A.~Huang, A.~Guez,
  T.~Hubert, L.~Baker, M.~Lai, A.~Bolton \emph{et~al.}, ``{Mastering the Game
  of Go without Human Knowledge},'' \emph{Nature}, vol. 550, no. 7676, pp.
  354--359, 2017.

\bibitem{openaiFive}
\url{https://openai.com/projects/five/} accessed on 01.07.2022.

\bibitem{TeslaPytorch}
A.~Karpathy, ``{Pytorch at Tesla},'' in \emph{{Pytorch DevCon'19}}, 2019.

\bibitem{mao2016resource}
H.~Mao, M.~Alizadeh, I.~Menache, and S.~Kandula, ``Resource management with
  deep reinforcement learning,'' in \emph{ACM HotNets}, 2016, pp. 50--56.

\bibitem{valadarsky2017learning}
A.~Valadarsky, M.~Schapira, D.~Shahaf, and A.~Tamar, ``Learning to route,'' in
  \emph{ACM HotNets}, 2017.

\bibitem{IFIPnetworking}
O.~{Iacoboaiea}, J.~{Krolikowski}, Z.~{Ben Houidi}, and D.~{Rossi},
  ``{Real-Time Channel Management in WLANs: Deep Reinforcement Learning versus
  Heuristics},'' in \emph{IFIP Networking}, 2021.

\bibitem{Vesselinova_2020}
N.~Vesselinova, R.~Steinert, D.~F. Perez-Ramirez, and M.~Boman, ``Learning
  combinatorial optimization on graphs: A survey with applications to
  networking,'' \emph{IEEE Access}, 2020.

\bibitem{zsmieeenet20}
C.~Benzaid and T.~Taleb, ``{AI}-driven zero touch network and service
  management in {5G} and beyond: Challenges and research directions,''
  \emph{IEEE Network}, vol.~34, no.~2, pp. 186--194, 2020.

\bibitem{zsm-wlan22}
M.~Friesen, L.~Wisniewski, and J.~Jasperneite, ``Machine learning for
  zero-touch management in heterogeneous industrial networks -- a review,'' in
  \emph{IEEE International Conference on Factory Communication Systems (WFCS)},
  2022.

\bibitem{AWSdeepracer}
\url{https://aws.amazon.com/deepracer} accessed on 1.07.2022.

\bibitem{TurboCA}
A.~Bhartia, B.~Chen, F.~Wang, D.~Pallas, R.~Musaloiu-E, T.~T.-T. Lai, and
  H.~Ma, ``{Measurement-Based, Practical Techniques to Improve 802.11ac
  Performance},'' in \emph{ACM IMC}, 2017.

\bibitem{Keshav}
N.~{Ahmed} and S.~{Keshav}, ``{A Successive Refinement Approach to Wireless
  Infrastructure Network Deployment},'' in \emph{IEEE WCNC}, 2006.

\bibitem{power_limitations}
V.~Shrivastava, D.~Agrawal, A.~Mishra, S.~Banerjee, and T.~Nadeem,
  ``{Understanding the Limitations of Transmit Power Control for Indoor
  {WLANs}},'' in \emph{ACM IMC}, 2007.

\bibitem{Sutton1998}
R.~S. Sutton and A.~G. Barto, \emph{{Reinforcement Learning: An Introduction}},
  2nd~ed.\hskip 1em plus 0.5em minus 0.4em\relax The MIT Press, 2018.

\end{thebibliography}

\pagebreak

\section*{Biographies}\label{sec:biographies}

%\begin{biography} 	
\textbf{Ovidiu-Constantin Iacoboaiea} is a Senior Research Engineer at the DataCom Lab of Huawei’s Paris Research Center. He received his MSc from the University Politehnica of Bucharest (UPB, Romania) and École supérieure d’électricité (Supélec, France), in 2012, and his PhD degree in Self-organizing networks from Telecom ParisTech (France) working with Orange Labs in 2015. Afterwards, he worked for Bouygues Telecom until 2019 as a System Engineer responsible for mobile network analysis, modeling, optimization and dimensioning. His current interests include network optimization, artificial intelligence, machine learning and big data. 
%\end{biography}

%\begin{biography} 
\textbf{Jonatan Krolikowski} is a Senior Research Engineer at the DataCom Lab of Huawei’s Paris Research Center. He received his MSc in Mathematics from TU Berlin in 2014 and his PhD in 2018 from Université Paris-Sud/CentraleSupélec/L2S on the topic of Optimal Content Management and Dimensioning in Wireless Networks. His current research interests are in network optimization applied to real networks leveraging methods from machine learning and operations research, as well as modelling and analysis of network related problems. %\end{biography}

% must be 50-100 words!!
%\begin{biography} 	
\textbf{Zied Ben Houidi} is a Principal Engineer at the DataCom Lab of the Paris Research Center. He received his Ph.D. from the University of Pierre et Marie Curie in France in December 2010 while he was working at Orange Labs on data-driven performance analysis of core networks’ routing protocols. He then joined Bell Labs, the research arm of Nokia. There, he proposed and led various research projects on network data valorization (e.g. human-level behavior analytics, learning from network/ISP data to build recommender systems) as well as automated reasoning for programmatic specifications/standards generation. The projects led to several deployments, patents and demos as well as publications in top tier competitive venues in Networking (e.g. ACM HotNets) and Social Computing (ACM CSCW/Proc. ACM on CHI, ACM Trans. on the Web). %\end{biography}

% must be 50-100 words!!
%\begin{biography} 
\textbf{Dario Rossi} is a Chief Expert on Network AI and Measurement, and director of the DataCom Lab of Huawei’s Paris Research Center. Previously, he was Chair professor at the Computer Science department of Telecom ParisTech (2006-2018) and Professor at Ecole Polytechnique (2012-2018). He received his MSc and PhD degrees in from Politecnico di Torino in 2001 and 2005 respectively, and was a visiting researcher at University of California, Berkeley during 2003-2004. He has coauthored 10+ patents and 150+ conference/journal papers on different aspects of networking, with 6000+ citations and an H-index of 40. He received several best paper awards, a Google Faculty Research Award (2015) and an IRTF Applied Network Research Prize (2016). He is a Senior Member of IEEE and ACM. 
%\end{biography}

\end{document}